\documentclass[a4paper,11pt]{article}
\pdfoutput=1 % if your are submitting a pdflatex (i.e. if you have
\usepackage{jheppub,wrapfig} % for details on the use of the package, please
\usepackage[utf8]{inputenc}
\usepackage{verbatim,xcolor}

\newcommand{\Vg}{V^\mu{}_\nu}
\newcommand{\Vf}{\widetilde{V}^\mu{}_\nu}
\newcommand{\Gg}{G^\mu{}_\nu}
\newcommand{\Gf}{\widetilde{G}^\mu{}_\nu}
\newcommand{\Tg}{T^\mu{}_\nu}
\newcommand{\Tf}{\widetilde{T}^\mu{}_\nu}
\newcommand{\Teff}{\mathcal{T}^\mu{}_\nu}
\newcommand{\fTeff}{\widetilde{\mathcal{T}}^\mu{}_\nu}
\newcommand{\Id}{\delta^\mu{}_\nu}

\newcommand{\Sqrt}{S^\mu{}_\nu}
\newcommand\tr{\mathsf{{\scriptscriptstyle T}}}
\newcommand{\Tgfluid}{T_\mathrm{fluid}{}^\mu{}_\nu}
\newcommand{\mv}{m_g}
\newcommand{\mgconst}{\widehat{m}_g}
\newcommand{\mfconst}{\widehat{m}_f}
\newcommand{\Qv}{Q_g}
\newcommand{\Qconst}{\widehat{Q}_g}

\global\long\def\change#1{\textcolor{black}{#1}}

\global\long\def\shiftpol#1#2#3{\left< #1 \right>_#2^#3}

\title{\boldmath \change{Generalized Vaidya solutions in bimetric gravity}}

\author{Marcus Högås,}
\author{Mikica Kocic,}
\author{Francesco Torsello and}
\author{Edvard Mörtsell}

\affiliation{The Oskar Klein Centre, Department of Physics, Stockholm University, SE 106 91, Stockholm, Sweden}

\emailAdd{marcus.hogas@fysik.su.se}
\emailAdd{mikica.kocic@fysik.su.se}
\emailAdd{francesco.torsello@fysik.su.se}
\emailAdd{edvard@fysik.su.se}

\abstract{In general relativity, the endpoint of spherically symmetric gravitational collapse is a Schwarzschild--[(A)dS] black hole. In bimetric gravity, it has been speculated that a static end state must also be Schwarzschild--[(A)dS]. To this end, we present a set of exact solutions, including collapsing massless dust particles. For these, the speculation is confirmed.}

\begin{document}
\maketitle
\flushbottom

\section{Introduction and background}
The dark sectors of the Universe suggest that there should be physics beyond general relativity (GR) or the standard model of particle physics. This provides an impetus to look for new theories of gravity beyond Einstein's. The problem of quantizing gravity may serve as additional motivation for that search.

Almost fifty years ago, Lovelock provided a road map, showing how modified theories can be constructed by systematically breaking the assumptions of GR \cite{Lovelock:1971yv}. Two such possibilities are to add new field content and to make gravity massive. Both alternatives have a long history; adding new fields with novel couplings to the metric dates back at least to Brans and Dicke \cite{Brans:1961sx}. A theory of massive gravity was first written down by Fierz and Pauli, who constructed a linearized theory for a massive spin-2 field propagating on a flat background \cite{Fierz:1939ix}. Three decades later, van Dam, Veltman, and Zakharov showed that the theory does not converge to GR in the zero-mass limit \cite{vanDam:1970vg,Zakharov:1970cc}, observationally disfavoring massive gravity. However, shortly thereafter Vainshtein found that the linearized theory becomes invalid below a certain distance scale, the Vainshtein radius, where a nonlinear completion needs to be considered \cite{Vainshtein:1972sx}. At the same time, Boulware and Deser showed that such a completion generically results in a ghost mode \cite{Boulware:1973my}. Therefore, in the following four decades there was little progress in the field. More recently, this no-go statement has been circumvented, leading to the construction of consistent theories of massive gravity \cite{deRham:2010ik,deRham:2010kj,Hassan:2011hr}. Bimetric theory is such a theory as it exhibits massive modes in addition to massless ones \cite{Hassan:2012wr}. For the construction of these massive modes, a second metric must be introduced and hence there is also additional field content. For recent reviews, see \cite{Schmidt-May:2015vnx,deRham:2014zqa}.

\noindent Bimetric theory is consistent insofar as it avoids the Boulware--Deser ghost \cite{Hassan:2011zd,Hassan:2011ea,Hassan:2018mbl}, and its unambiguous space-time interpretation was established in \cite{Hassan:2017ugh}. To be viable, the theory should also exhibit solutions which agree with observations. Therefore, it is crucial to analyze the spectrum of solutions to the equations of motion. Ultimately, this requires full-fledged numerical integration of the nonlinear partial differential equations \eqref{eq:EoM} in four space-time dimensions, including realistic matter sources. Even more so in bimetric theory than GR; due to the Vainshtein mechanism invalidating linear perturbation theory under some circumstances, the only options left in those regions are exact, nonlinear solutions and numerical simulations. Although some steps have been taken towards numerical bimetric relativity \cite{Kocic:2018ddp,Kocic:2018yvr,cbssn,Kocic:2019zdy,polytropes,meangauge,bimEX}, the space of dynamical solutions is largely unknown. On the other hand, under the assumption of symmetries, exact closed-form solutions can be found. Their simplicity makes them serve as useful toy models, approximating more realistic setups. 

To this end, the bimetric equations of motion have been solved under different assumptions. Cosmological models assume homogeneous and isotropic metrics. These solutions can reproduce the accelerated expansion of the Universe, without introducing a cosmological constant, see for example \cite{vonStrauss:2011mq,Volkov:2011an,Comelli:2011zm,Akrami:2012vf,Akrami:2013pna,Nersisyan:2015oha,Dhawan:2017leu}. \change{They are also compatible with local tests of gravity and gravitational wave observations, see \cite{Luben:2018ekw}.}

In vacuum, a solution to Einstein's equations (with or without cosmological constant) is also solution to the bimetric equations of motion \cite{Kocic:2017wwf,Hassan:2014vja}. Assuming spherical symmetry as well, the GR solution is Schwarzschild--[(A)dS] due to Birkhoff's theorem \cite{Jebsen:1921,Jebsen2005,Birkhoff:1923,Kocic:2017hve}.

Nevertheless, there are static, spherically symmetric solutions of the bimetric equations which do not solve Einstein's equations \cite{Babichev:2013pfa,Volkov:2012wp}. However, these exhibit asymptotic behavior diverging from Minkowski and (A)dS \cite{Torsello:2017cmz}. On the other hand, in the presence of matter sources, there are non-GR solutions approaching Minkowski or (A)dS when $r \to \infty$ \cite{Enander:2015kda}. Imagining one of these matter solutions as the initial state of gravitational collapse and that one of the non-GR vacuum solutions is the end state, the asymptotic structure must change during collapse, which we do not expect to happen. Therefore, it has been speculated that if gravitational collapse ends in a static state, it must be a GR solution, that is, a Schwarzschild--[(A)dS] black hole. The outstanding question at this point is which ones (if any) of the static, spherically symmetric black holes that represent possible end states of gravitational collapse. For the class of solutions presented in this paper, indeed the end states are GR black holes.

\paragraph{Summary of results.} A set of non-static, spherically symmetric solutions for the two metrics, $g$ and $f$, in the presence of a generalized Vaidya fluid \cite{Wang:1998qx}, is:
\begin{subequations}
	\label{eq:solution}
	\begin{align}
	g&= -\left(1- \frac{2 m(v,r)}{r}\right) \mathrm{d}v^2 + 2 \mathrm{d}v \mathrm{d}r +r^2 \mathrm{d}\Omega^2,\\
	f &= c^2 \Bigg[-\left(1-\frac{2 \widetilde{m}(v,r)}{r}\right) \mathrm{d}v^2 + 2 \mathrm{d}v \mathrm{d}r + r^2 \mathrm{d}\Omega^2 \Bigg],
	\end{align}
\end{subequations}
where $m$ is the Misner--Sharp mass function of $g$, $\widetilde{m}$ is the Misner--Sharp mass of $f$, and $c$ is a constant. The stress--energy of the generalized Vaidya fluid, coupled to $g$, is generated from the Einstein tensor of $g$ and the bimetric stress--energy which arises from the interaction between the metrics. The solutions can be divided into two branches.

In Branch I, $m$ is arbitrary up to some basic conditions, possibly implied by the weak, strong, and dominant energy conditions \cite{Wang:1998qx}. Letting,
\begin{equation}
m(v,r) = \mv(v) - \frac{\Qv^2(v)}{2r} + \frac{\lambda}{6}r^3,
\end{equation}
yields the stress--energy of a collapsing cloud of massless, electrically charged particles in a cosmological (A)dS background, with $\mv(v)$ describing the black hole mass as a function of advanced time, $v$, and $\Qv(v)$ the electric charge. Concerning the Misner--Sharp mass of $f$, $\widetilde{m}(v,r) = \mfconst + \widetilde{\Lambda} r^3 /6$, with $\mfconst$ being the constant black hole mass in the $f$-sector and $\widetilde{\Lambda}$ is a constant given in terms of the $\beta$-parameters. The $f$ metric is therefore a Schwarzschild--[(A)dS] metric in Eddington--Finkelstein coordinates. Generically, it exhibits a curvature singularity at $r=0$, covered by the event horizon with respect to the same metric, but only partially covered by the event horizon of $g$, see figure \ref{fig:br-vaidya-horizons-and-mass}. This calls a bimetric analog of the cosmic censorship hypothesis into question. However, the naked singularity can be eliminated by choosing the mass of the $f$-sector black hole to be zero, reducing $f$ to (A)dS/Minkowski. At the end state of collapse, $\mv(v)$ and $\Qv(v)$ are constants and $g$ is Reissner--Nordström--[(A)dS], recovering the non-bidiagonal static, spherically symmetric GR solutions of \cite{Comelli:2011wq,Babichev:2014fka}.

In Branch II,
\begin{equation}
\label{eq:mBranchIIa}
	m(v,r) =   \mv(v) + \frac{1}{6} \widetilde{\Lambda} c^2 r^3+ 2 \frac{\ell^2}{\kappa_g} \frac{c}{\beta_1 + 2 \beta_2 c + \beta_3 c^2} \frac{\partial_v \; \mv(v)}{r},
\end{equation}
generating the stress--energy of an effective cosmological constant plus a traceless, Type II \cite{Hawking:1973uf}, fluid with equation of state $w = P/\rho = 1$ (an ultrastiff fluid \cite{rezzolla2013relativistic}). Here, $c$ is a free constant. The $f$-sector mass function takes the same form as $m$ except the $1/r$ term, which is missing. For $\beta_1 + 2 \beta_2 c + \beta_3 c^2<0$, corresponding to an imaginary Fierz--Pauli mass \change{in Branch II}, the energy conditions can always be satisfied for appropriate choices of $\mv(v)$, whereas if the Fierz--Pauli mass is real, they are always violated outside some critical radius. However, with the latter choice, there is no gravitational collapse, but a radiating solution. In the former case, if $\ell$ is not too big, the singularity formed by the collapse is covered by event horizons with respect to both metrics. The static end states of this branch are the well-known proportional ($f = c^2 g$), static and spherically symmetric vacuum solutions. 

\paragraph{Conventions.} Tildes denote quantities constructed from $f$, otherwise constructed from $g$. For example, $\nabla$ is the covariant derivative with respect to $g$, and $\widetilde{\nabla}$ is the covariant derivative with respect to $f$. Occasionally, we attach labels $g$ and $f$, for example, $\mgconst$ is the mass parameter of a Schwarzschild metric in the $g$-sector. The metrics are written in ingoing Eddington--Finkelstein-like coordinates, with $v$ being the advanced time. Usually, $v$ is referred to as ``time". Geometrized units are employed, in which Newton's gravitational constant and the speed of light are set to one. \newpage

\subsection{Generalized Vaidya solutions in general relativity}
\label{sec:Tobs}
The Schwarzschild metric in ingoing Eddington--Finkelstein coordinates $(v,r,\theta,\phi)$, is, 
\begin{equation}
\label{eq:GRVaidya}
	g = -\left(1-\frac{2m}{r}\right)\mathrm{d}v^2 + 2 \mathrm{d}v\mathrm{d}r +r^2 \mathrm{d}\Omega^2, \quad \mathrm{d}\Omega^2 := \mathrm{d}\theta^2 + \sin^2 \theta \, \mathrm{d}\phi^2,
\end{equation}
with $m =\mathrm{const.}$ Including a dependence of $m$ on $v$ and $r$, yields the generalized Vaidya metric \cite{Vaidya:1951zz,Wang:1998qx}. With this dependence, the Einstein tensor is non-zero and a stress--energy must be introduced. Einstein's equations are:
\begin{equation}
	\Gg = \kappa_g \Tg.
\end{equation}
In this setup, we define $\Tg$ to be whatever we obtain when calculating $\Gg$. This procedure is sometimes referred to as Synge's method and occasionally the stress--energy defined this way can be generated by a reasonable matter model. A choice of $m$, satisfying this demand, is,
\begin{equation}
\label{eq:dustchargecosm}
m(v,r) = \mv(v) - \frac{\Qv^2(v)}{2r} + \frac{\lambda}{6}r^3 .
\end{equation}
The function $\mv(v)$ is interpreted as the mass of the black hole, with the corresponding stress--energy generated by infalling massless particles. The $1/r$ term represents the electric charge, $\Qv(v)$, of the black hole, with the stress--energy being generated by an electric field \cite{Bonnor:1970zz,Ori}. Finally, the $r^{3}$ term is a cosmological constant contribution. To satisfy the weak, strong, and dominant energy conditions, \eqref{eq:energycond} must hold, imposing some conditions on $m$. Neutrinos, being very light and weakly interacting, fits a model with $\Qv=0$ well \cite{Lindquist:1965zz}. With $\Qv=\lambda=0$, starting with $m=0$ at $v=0$, letting $m$ grow smoothly in the interval $0 \leq v \leq v_f$ (corresponding to infall of massless dust particles), the space-time geometry goes through a transition, starting with Minkowski, transiting through an intermediate Vaidya phase, and ending up in a Schwarzschild black hole of mass $m(v_f) = \mathrm{const}$. In summary, the generalized Vaidya metric \eqref{eq:GRVaidya} with $m$ given by \eqref{eq:dustchargecosm} gives the metric of a spherically symmetric distribution of massless particles carrying some electrical charge, collapsing in a cosmological (A)dS background. There are also other forms of $m(v,r)$ generating physical matter fields, see for example \cite{Barriola:1989hx,Husain:1995bf}.

With $m(v \leq 0)=0$ and $\mv(v)$ and $\Qv(v)$ growing in the time interval $0 \leq v \leq v_f$, a calculation of the Kretschmann scalar reveals that there is a curvature singularity at $r=0$ when  $v \geq 0$. Generically, this singularity is covered by an event horizon. Nevertheless, there are space-times where the singularity is not covered completely, in which case there is a naked singularity \cite{Papapetrou}. However, the pathology can be blamed on the matter model, which becomes invalid in regions of very high curvature. 

\paragraph{Stress--energy.} The stress--energy for a generalized Vaidya metric can be decomposed as (see, e.g., \cite{Wang:1998qx}),
\begin{equation}
\label{eq:stressenergydecomp}
T_{\mu\nu} = \mu l_\mu l_\nu + 2 (\rho + P) l_{(\mu} n_{\nu)} + P g_{\mu\nu},
\end{equation}
where we have introduced the ingoing and outgoing null covectors, respectively,
\begin{equation}
\label{eq:RNGout}
l_\mu := -\partial_\mu v, \quad n_\mu := -\frac{1}{2}\left(1- \frac{2m}{r}\right) \partial_\mu v + \partial_\mu r .
\end{equation}
Here, $\rho$ is the total energy density, $\mu$ is the total energy flux in the $l^\mu$ direction, and $P$ is the total tangential (principal) pressure.
\begin{equation}
\label{eq:Tgeigenval}
\rho = - T^v {}_v= 2 \frac{\partial_r m}{r^2}, \quad  \mu = T^r{}_v =  2 \frac{\partial_v m}{r^2}, \quad P = T^\theta{}_\theta = - \frac{\partial_r^2 m}{r}.
\end{equation}
The stress--energy is of Type II in the classification of \cite{Hawking:1973uf}. For such a fluid, the dominant energy condition reads,
\begin{equation}
\label{eq:energycond}
\mu \geq 0, \quad \rho \geq P \geq 0, \quad (\mu \neq 0).
\end{equation}
If these inequalities hold, the weak and strong energy conditions are satisfied as well. The dominant energy condition guarantees that the energy flow of the fluid does not exceed the speed of light. It should be stressed that the energy conditions present neither necessary nor sufficient requirements for a solution to be regarded as physical, they rather serve as a useful diagnostics of the solution.

\subsection{Bimetric gravity}
Bimetric theory is a nonlinear theory of two interacting, symmetric, rank-2 tensor fields (i.e., metrics) defined on the same manifold. The Hassan--Rosen action is, in natural units,
\begin{equation}
\label{eq:action}
\mathcal{S}_\mathrm{HR} = \int d^4 x \left[\frac{1}{2\kappa_g}  \sqrt{-g} R + \frac{1}{2\kappa_f} \sqrt{-f} \widetilde{R} - \frac{1}{\ell^2} \sqrt{-g} \sum_{n=0}^{4} \beta_n e_n(S) + \sqrt{-g} \mathcal{L}_\mathrm{m} + \sqrt{-f} \widetilde{\mathcal{L}}_\mathrm{m}\right],
\end{equation}
where $e_n(S)$ are the elementary symmetric polynomials of the principal square root $S := (g^{-1} f)^{1/2}$ \cite{Hassan:2011zd}. The $\beta$-parameters are dimensionless constants, to be fixed by observations. Since $e_0(S)=1$, the $\beta_0$ term contributes to the action as a cosmological constant with respect to $g$. Similarly, the $\beta_4$ term provides a cosmological constant in the $f$-sector.

Generically, there can be two independent sectors of matter fields, contained in the matter Lagrangians $\mathcal{L}_\mathrm{m}$ and $\widetilde{\mathcal{L}}_\mathrm{m}$, minimally coupled to $g$ and $f$ respectively \cite{deRham:2014fha,deRham:2014naa}. The $g$ metric determines the geodesics of the $g$-sector matter. Thus, $g$ can be identified as the physical metric with respect to $g$-sector observers. Such an observer can only measure the geometry of $g$ directly; the $f$-sector influences the observers only indirectly via its interaction with the physical metric.

Varying \eqref{eq:action} with respect to $g$ and $f$ yields two copies of Einstein's equations with effective stress--energies $\Teff$ and $\fTeff$,
\begin{subequations}
	\label{eq:EoM}
	\begin{align}
	\label{eq:gEoM}
	&\Gg = \kappa_g \Teff, \qquad \Teff:= \Tg + \Vg ,\\
	\label{eq:fEoM}
		&\Gf =  \kappa_f \fTeff, \qquad \fTeff := \Tf + \Vf,
	\end{align}
\end{subequations}
where $\Tg$ and $\Tf$ are the ordinary matter stress--energies coupled to $g$ and $f$ respectively. The bimetric stress--energy tensors, ${V^\mu}_\nu$ and $\Vf$, contain the interaction between the two metrics and are defined as,
\begin{subequations}
	\begin{align}
	\Vg &:= -\frac{1}{\ell^2} \sum_{n=0}^{3} \beta_n \sum_{k=0}^{n} (-1)^{n+k} e_k(S) {(S^{n-k})}^\mu{}_\nu,\\
	\Vf &:= - \frac{1}{\ell^2} \sum_{n=0}^{3}\beta_{4-n} \sum_{k=0}^{n} (-1)^{n+k} e_k(S^{-1}) {(S^{-n+k})}^\mu{}_\nu.
	\end{align}
\end{subequations}
The bimetric conservation law reads,
\begin{equation}
\label{eq:Bianchi}
\nabla_\rho {V^\rho}_\mu = 0.
\end{equation}

\section{Generalized Vaidya solutions in bimetric gravity}
\change{As in general relativity, there are several motivations for the study of generalized Vaidya solutions in bimetric gravity. For example, they approximate the gravitational collapse of light and weakly interacting particles like neutrinos, and serve as useful toy models in the study of the cosmic censorship hypothesis. Further, the solutions presented in this paper are the only exact solution of gravitational collapse in bimetric gravity as there are no Lemaître--Tolman--Bondi solutions, except homogeneous and isotropic ones which constitute a special case \cite{Hogas:2019ywm}.}

\subsection{Ansatz}
Assuming that $g$ and $f$ exhibit the same spherical symmetry \cite{Torsello:2017ouh} and that the two metrics have a common null direction, the most general Ansatz is,
\begin{subequations}
	\label{eq:genAnsatz}
	\begin{align}
	g &= -e^{2p(v,r)} \left(1- \frac{2 m(v,r)}{r}\right) \mathrm{d}v^2 + 2 e^{p(v,r)} \mathrm{d}v \mathrm{d}r +r^2 \mathrm{d}\Omega^2,\\
	f &= c^2(v,r) \Bigg[-e^{2q(v,r)}\left(1-\frac{2 \widetilde{m}(v,r)}{r}\right) \mathrm{d}v^2 + 2 e^{q(v,r)}\mathrm{d}v \mathrm{d}r + r^2 \mathrm{d}\Omega^2 \Bigg].
	\end{align}
\end{subequations}
The hypersurface defined by $v= \mathrm{const.}$ is a null surface with respect to both metrics. To obtain a closed-form solution, we further restrict the Ansatz \eqref{eq:genAnsatz}, demanding that $p = q = 0$ and $c = \mathrm{const}$. Hence,
\begin{subequations}
	\label{eq:Ansatz}
	\begin{align}
	\label{eq:Ansatzg}
	g &= -\left(1- \frac{2 m(v,r)}{r}\right) \mathrm{d}v^2 + 2 \mathrm{d}v \mathrm{d}r +r^2 \mathrm{d}\Omega^2,\\
	\label{eq:Ansatzf}
	f &= c^2 \Bigg[-\left(1-\frac{2 \widetilde{m}(v,r)}{r}\right) \mathrm{d}v^2 + 2 \mathrm{d}v \mathrm{d}r + r^2 \mathrm{d}\Omega^2\Bigg].
	\end{align}
\end{subequations}
The metrics are of the generalized Vaidya type, with $m$ being the Misner--Sharp mass of $g$ \cite{Misner:1964je} (also referred to as Hernandez--Misner mass, and  coinciding with the Hawking--Israel mass in spherical symmetry) and $\widetilde{m}$ the Misner--Sharp mass of $f$, defined by, respectively,
\begin{equation}
	m_\mathrm{MS} := \frac{r}{2} \big[1-g^{\mu\nu} \left(\nabla_\mu r
	\right) \left(\nabla_\nu r\right)\big] = m, \quad \widetilde{m}_\mathrm{MS} := \frac{r}{2} \big[1-f^{\mu\nu} (\widetilde{\nabla}_\mu r) (\widetilde{\nabla}_\nu r)\big] = \widetilde{m}\nonumber.
\end{equation}
An Ansatz similar to \eqref{eq:Ansatz} was adopted in \cite{Kocic:2017hve}, the difference being that here we solve the equations of motion (EoM) in the presence of a generalized Vaidya fluid generated from $\Gg$ and $\Vg$ and that we do not restrict our space of $\beta$-parameters. With the Ansatz \eqref{eq:Ansatz}, the square root is,
\begin{equation}
\label{eq:S}
	{S^\mu}_\nu = |c| \left(\begin{array}{cccc}
	1 & 0 & 0 & 0\\
	\left[\widetilde{m}(v,r)-m(v,r)\right]/r & 1 & 0 & 0\\
	0 & 0 & 1 & 0\\
	0 & 0 & 0 & 1
	\end{array}\right).
\end{equation}
Due to the ubiquity of the $|c|$'s, we hereafter let the modulus be understood. Note that ${S^\mu}_\nu$ contains a factor $\widetilde{m}-m$, reappearing in the bimetric stress--energies ${V^\mu}_\nu(S)$ and $\Vf(S)$. In other words, there is dynamical interaction between the metrics, at least off-shell. Writing ${S^\mu}_\nu$ in Jordan normal form reveals that it is of Type IIa if $m \neq \widetilde{m}$ and of Type I if $m = \widetilde{m}$, in the classification of Hassan and Kocic \cite{Hassan:2017ugh}. The traces of powers of the square root are,
\begin{equation}
\label{eq:Scoordinv}
	\mathrm{Tr}[S] = 4|c|, \;\; \mathrm{Tr}[S^2] = 4|c|^2, \;\; \mathrm{Tr}[S^3] = 4|c|^3, \;\; \mathrm{Tr}[S^4] = 4|c|^4.
\end{equation}
Thus, within the coordinate range, there is no singularity in the interaction terms as long as $0 < |c| < \infty$, which we assume. In section \ref{sec:singandhor}, we discuss the behavior of $S$ at $r=0$.

\subsection{Equations of motion and solutions}
\label{sec:EoMsol}
With the Ansatz \eqref{eq:Ansatz}, the Einstein tensors read,
\begin{subequations}
	\label{eq:tensorG}
	\begin{alignat}{2}
	\label{eq:Gg}
		\Gg &=& \frac{1}{r^2} \times &\left(\begin{array}{cccc}
		-2 \partial_r m & 0 & 0& 0\\
		2\partial_v m & -2 \partial_r m  & 0 & 0\\
		0 & 0 & - r \partial_r^2 m & 0\\
		0 & 0 & 0 & - r \partial_r^2 m
		\end{array}\right),\\
	\label{eq:Gf}
		\Gf &=& \frac{1}{c^2 r^2} \times &\left(\begin{array}{cccc}
		-2 \partial_r \widetilde{m} & 0 & 0& 0\\
		2\partial_v \widetilde{m} & -2 \partial_r \widetilde{m}  & 0 & 0\\
		0 & 0 & - r \partial_r^2\widetilde{m} & 0\\
		0 & 0 & 0 & - r \partial_r^2 \widetilde{m}
		\end{array}\right),
	\end{alignat}
\end{subequations}
and the bimetric stress--energy tensors,
\begin{subequations}
	\label{eq:tensorV}
	\begin{alignat}{2}
	\label{eq:tensorVg}
	{V^\mu}_\nu &= - \frac{1}{\kappa_g} \left(\begin{array}{cccc}
	\Lambda & 0 & 0 & 0\\
	\kappa_g c \shiftpol{c}{1}{2} \dfrac{m-\widetilde{m}}{r} & \Lambda & 0 & 0\\
	0 & 0 &\Lambda & 0\\
	0 & 0 & 0 & \Lambda
	\end{array}\right),& \quad \Lambda &:= \kappa_g \shiftpol{c}{0}{3},\\
	\label{eq:tensorVf}
	\Vf &= - \frac{1}{\kappa_f}\left(\begin{array}{cccc}
	\widetilde{\Lambda} & 0 & 0 & 0\\
	\kappa_g \dfrac{\shiftpol{c}{1}{2}}{c^3} \dfrac{\widetilde{m}-m}{r} & \widetilde{\Lambda} & 0 & 0\\
	0 & 0 & \widetilde{\Lambda} & 0\\
	0 & 0 & 0 &\widetilde{\Lambda}
	\end{array}\right),& \quad \widetilde{\Lambda} &:=  \change{\kappa_f \frac{\shiftpol{c}{1}{3}}{c^3}},
	\end{alignat}
\end{subequations}
with the shifted elementary symmetric polynomials defined as,
\begin{equation}
	\shiftpol{X}{k}{n} := - \frac{1}{\ell^2} \sum_{i=0}^{n} \binom{n}{i} \beta_{i+k} X^i.
\end{equation}
From \eqref{eq:tensorVg} we compute the bimetric conservation law \eqref{eq:Bianchi}:
\begin{equation}
\label{eq:Bianchicalc}
	\left(\beta_1 + 2 \beta_2 c + \beta_3 c^2 \right) \left[(m - \widetilde{m}) - r \partial_r  \left( m - \widetilde{m}\right) \right] = 0.
\end{equation}
There are two branches of solutions: setting the first parenthesis to zero (Branch I) and setting the second one to zero (Branch II).

\paragraph{Branch I.} Setting the first parenthesis to zero,
\begin{subequations}
	\label{eq:cBranchI}
	\begin{alignat}{2}
		c &= \left(-\beta_2 \pm \sqrt{\beta_2^2 -\beta_1 \beta_3 }\right) / \beta_3, & \quad \beta_3 &\neq 0,\\
		c &= - \beta_1 / \left(2 \beta_2\right), &\beta_3 &= 0.
	\end{alignat}
\end{subequations}
Note that this choice is different from setting the Fierz--Pauli mass to zero since $g$ and $f$ are not proportional \change{(see appendix \ref{sec:FPmass} for a detailed discussion)}. The off-diagonal components of $\Vg$ and $\Vf$ are proportional to $\beta_1 + 2 \beta_2 c + \beta_3 c^2$, and hence vanish and the only remaining components are the (constant) diagonal ones, that is,
\begin{equation}
\label{eq:Vsol}
\Vg = - \kappa_g^{-1} \Lambda \: \Id, \;\; \Vf =- \kappa_f^{-1} \widetilde{\Lambda} \: \Id. 
\end{equation}
At this point, the contribution from the bimetric interaction has reduced to an effective cosmological constant in each sector: $\Gg +\Lambda \Id= \kappa_g \Tg$ and $\Gf +\widetilde{\Lambda} \Id=\kappa_f \Tf$. From the definitions of $\Lambda$ and $\widetilde{\Lambda}$ \eqref{eq:tensorV} it is apparent that they involve all the $\beta$-parameters and arise because of bimetric interaction.

Assuming $\Tf=0$, the $f$-sector EoM yield two differential equations for $\widetilde{m}$ which are readily solved, with the result,
\begin{equation}
\label{eq:mtildeBranchI}
	\widetilde{m}(v,r) = \mfconst + \frac{\widetilde{\Lambda}}{6} c^2 r^3,
\end{equation}
where $\mfconst$ is an integration constant. The $f$ metric \eqref{eq:Ansatzf} with \eqref{eq:mtildeBranchI} is a Schwarzschild--[(A)dS] black hole with mass $\mfconst$ and cosmological constant $\widetilde{\Lambda}c^2$. Unless $\mfconst$ is set to zero, there is a curvature singularity at $r=0$, covered by an event horizon with respect to $f$.

To solve the EoM in the $g$-sector \eqref{eq:gEoM} we apply Synge's method and define $\Tg$ to be, 
\begin{equation}
\label{eq:stressenergy}
	\Tg := \kappa_g^{-1} \Gg - \Vg.
\end{equation}
Since $\Vg$ is an effective cosmological constant \eqref{eq:Vsol}, the generalized Vaidya metric, \eqref{eq:Ansatzg}, with $m$ of the form \eqref{eq:dustchargecosm}, is as physical as in GR. For example, to generate the stress--energy of a collapsing cloud of massless dust particles in a cosmological background, $m(v,r) =\mv(v)+\Lambda r^3 /6$. The Branch I solutions can straightforwardly be generalized by including $\Tf$ in the $f$-sector EoM. In this case, both $g$ and $f$ are generalized Vaidya metrics. 

\noindent Concerning the stress--energy observables (see section \ref{sec:Tobs}), we split $\rho$, $\mu$, and $P$ into a cosmological constant contribution and a fluid contribution, for example the total energy density is $\rho = \rho_\mathrm{CC} + \rho_\mathrm{fluid}$. 
\begin{subequations}
\begin{alignat}{3}
	\rho_\mathrm{CC} &= -\Lambda, \quad  &\mu_\mathrm{CC} &= 0, \quad &P_\mathrm{CC} &= \Lambda,\\
	\rho_\mathrm{fluid} &= 2 \frac{\partial_r m}{r^2}, \quad  &\mu_\mathrm{fluid} &=   2 \frac{\partial_v m}{r^2}, \quad &P_\mathrm{fluid} &= - \frac{\partial_r^2 m}{r}.
\end{alignat}
\end{subequations}
The stress--energy tensor corresponding to the fluid is of Type II \cite{Hawking:1973uf}.

To summarize, the Branch I solutions are:
\begin{subequations}
	\label{eq:BranchI}
	\begin{align}
	\label{eq:BranchIg}
		m(v,r) &= \mv(v) - \frac{\Qv^2(v)}{2r} + \frac{\lambda}{6}r^3  ,\\
	\label{eq:BranchIf}
		\widetilde{m}(v,r) &=\mfconst + \frac{\widetilde{\Lambda}}{6} c^2 r^3,
	\end{align}
\end{subequations}
where we choose $m$ to yield massless, electrically charged particles collapsing in a cosmological (A)dS background. In general, the two metrics are not proportional and, moreover, cannot be diagonalized in the same coordinates. The isometry group of $g$ is SO(3) (spherical symmetry) whereas $f$ admits also a timelike Killing vector field and is thus static. With $\mfconst=0$, $f$ is maximally symmetric. Thus, the isometry groups of the two metrics do not coincide generically \cite{Torsello:2017ouh}.

\paragraph{Branch II.} Here, the bimetric conservation law \eqref{eq:Bianchicalc} is solved by setting the second parenthesis to zero. The differential equation is readily solved for $\widetilde{m}(v,r)$,
\begin{equation}
\label{eq:mtildeBranchII}
	\widetilde{m}(v,r) = m(v,r) + \frac{a(v)}{r},
\end{equation}
where $a(v)$ is a freely specifiable function of $v$. Plugging \eqref{eq:mtildeBranchII} into \eqref{eq:fEoM} with \eqref{eq:Gf}, \eqref{eq:tensorVf}, and $\Tf=0$ results in two differential equations. The solutions split into two cases, depending on whether $m$ depends on $r$. In Branch IIa, $\partial_r m \neq 0$, and,
\begin{subequations}
	\label{eq:BranchIIa}
	\begin{align}
	\label{eq:BranchIIag}
		m(v,r) &=   \mv(v) + \frac{1}{6} \widetilde{\Lambda} c^2 r^3 - \frac{2}{\kappa_g} \frac{c}{\shiftpol{c}{1}{2}} \frac{\partial_v \; \mv(v)}{r},\\
	\label{eq:BranchIIaf}
		\widetilde{m}(v,r) &= \mv(v) + \frac{1}{6} \widetilde{\Lambda} c^2 r^3,
	\end{align}
\end{subequations}
and in Branch IIb, $\partial_r m =0$, and,
\begin{equation}
	\label{eq:BranchIIb}
	m(v,r)= \widetilde{m}(v,r) =\mgconst,\quad \widetilde{\Lambda} = 0,
\end{equation}
where $\mgconst$ is a constant. In the latter branch, $g$ and $f$ are proportional Schwarzschild metrics. Concerning the $g$-sector EoM in Branch IIa, Synge's method is implemented again. The stress--energy can be generated by a special fluid plus an effective cosmological constant contribution:
\begin{equation}
	\Tg := \kappa_g^{-1} \Gg - \Vg = - \Lambda_\mathrm{eff} \Id + \Tgfluid,
\end{equation}
with (note the linearity in $\mv(v)$ and its derivatives),
\begin{subequations}
	\begin{align}
		\Lambda_\mathrm{eff} &:= c^2 \widetilde{\Lambda}-\Lambda,\\
		\Tgfluid &:= \frac{4}{\kappa_g^2} \frac{c}{\shiftpol{c}{1}{2}} \frac{\partial_v \mv(v)}{r^4} \left(\begin{array}{cccc}
		-1 & 0 &0 &0\\
		\frac{\kappa_g \shiftpol{c}{1}{2} (1+c^2)}{4c} r^2 - \frac{1}{2} r \, \frac{\partial_v^2 \mv(v)}{\partial_v \mv(v)} & -1 &0&0\\
		0&0&1&0\\
		0&0&0&1
		\end{array}\right).
	\end{align}
\end{subequations}
There is dynamical interaction on-shell. Note that $\Tgfluid$ is traceless and of Type II \cite{Hawking:1973uf}. Splitting the total stress--energy into the fluid contribution and the cosmological constant contribution,
\begin{subequations}
	\begin{alignat}{3}
	\rho_\mathrm{CC} &= -\Lambda_\mathrm{eff}, \quad &\mu_\mathrm{CC} &= 0, \quad &P_\mathrm{CC} &= \Lambda_\mathrm{eff},\\
	\label{eq:fluiddecomp}
	\rho_\mathrm{fluid} &= - T_\mathrm{fluid}{}^v{}_v, \quad &\mu_\mathrm{fluid} &= T_\mathrm{fluid}{}^r{}_v, \quad &P_\mathrm{fluid} &= T_\mathrm{fluid}{}^\theta{}_\theta.
	\end{alignat}
\end{subequations}
The equation of state of the fluid is,
\begin{equation}
	\label{eq:EoS}
	w_\mathrm{fluid} = P_\mathrm{fluid}/\rho_\mathrm{fluid} = 1,
\end{equation}
so it can be characterized as an ultrastiff fluid \cite{rezzolla2013relativistic}. Generalization of the solution by inclusion of $\Tf$ is not as straightforward as in Branch I.

In bimetric theory, with several stress--energies ($\Teff$, $\Tg$, $\Vg$, etc.), it is not obvious how the energy conditions should be used. This issue has only been discussed sporadically in the literature, see \cite{Capozziello:2014bqa,Baccetti:2012re}. For example, the bimetric stress--energy $\Vg$ usually violates the energy conditions for cosmological solutions \cite{Konnig:2015lfa}, and in vacuum the null energy conditions (NEC) with respect to $\Vg$ and $\Vf$ are strongly anti-correlated; if the NEC holds with respect to $\Vg$, it is violated for $\Vf$, and vice versa. Here we analyze the implications of imposing the dominant energy condition (implying also the weak and strong ones) on the stress--energy $\Tgfluid$. Imposing \eqref{eq:energycond},
\begin{subequations}
	\begin{alignat}{3}
		\label{eq:BIIaPhifluid}
		\mu_\mathrm{fluid} &\geq 0 \quad &&\Rightarrow \quad &\kappa_g r \, \partial_v \mv(v) &\geq \frac{2c}{\shiftpol{c}{1}{2}(1+c^2)} \, \partial_v^2 \mv(v) ,\\
		\label{eq:BIIaPfluid}
		P_\mathrm{fluid} &\geq 0 \quad &&\Rightarrow \quad &\frac{1}{\shiftpol{c}{1}{2}} \partial_v \mv(v) &\geq 0.
	\end{alignat}
\end{subequations}
The relation $\rho_\mathrm{fluid} \geq P_\mathrm{fluid}$ always holds, due to the equation of state \eqref{eq:EoS}. The conditions split into two cases, depending on the sign of $\shiftpol{c}{1}{2}$. For $\shiftpol{c}{1}{2} >0$, \eqref{eq:BIIaPfluid} is equivalent to,
\begin{equation}
\label{eq:mincr}
	\partial_v \mv(v) \geq 0,
\end{equation}
that is, $\mv(v)$ is monotonically increasing (or is constant). To satisfy \eqref{eq:BIIaPhifluid} for all $r$, 
\begin{equation}
\label{eq:dmdvdecr}
	\partial_v^2 \mv(v) \leq 0,
\end{equation}
otherwise there is a critical radius within which \eqref{eq:BIIaPhifluid} is violated. With \eqref{eq:mincr} and \eqref{eq:dmdvdecr}, $\partial_v \mv(v)$ is necessarily discontinuous at $v=0$, see figure \ref{fig:br-vaidya-horizons-and-mass-branchii}. For $\shiftpol{c}{1}{2} <0$, \eqref{eq:BIIaPfluid} is equivalent to,
\begin{equation}
\label{eq:dvmlesszero}
\partial_v \mv(v) \leq 0,
\end{equation}
so that $\mv(v)$ is monotonically decreasing (or constant). Thus, we must start with an initial black hole and then let it radiate. In GR, this is only possible when switching from advanced to retarded time. Moreover, due to \eqref{eq:dvmlesszero}, \eqref{eq:BIIaPhifluid} is necessarily violated outside some critical radius. If desired, this radius can be pushed arbitrarily far out by letting $M$ be sufficiently small. In the following, we assume that $\shiftpol{c}{1}{2}>0$, corresponding to an imaginary Fierz--Pauli mass around the proportional end state.

An alternative way of generating the stress--energy of Branch IIa is to look at \eqref{eq:BranchIIag} and identify the $1/r$ term as electric charge. Note however that this interpretation requires that the charge is fine-tuned to $\Qv^2(v) \propto \partial_v \mv(v)$.

In Branch IIa, the metrics are proportional if and only if $\partial_v \mv =0$, in which case they reduce to Schwarzschild--[(A)dS] metrics. If both metrics exhibit a time dependence, they have the same isometry group, SO(3), as well as the same Killing vector fields.

\subsection{Static end states}
\paragraph{Branch I.} In the static end state of gravitational collapse, $m(v,r)$ is of the form,
\begin{equation}
	m(v,r) = \mgconst - \frac{\Qconst^2}{2r} + \frac{\lambda}{6}r^3,
\end{equation}
with $\mgconst$ and $\Qconst$ being the (constant) final mass and charge of the $g$-sector black hole, and $\lambda$ being a cosmological constant. The end state of $g$ is a Reissner--Nordström--[(A)dS] metric, that is, an electrically charged black hole in a cosmological background or, if $\Qv=0$, a Schwarzschild--[(A)dS] metric. As for $f$, $\widetilde{m}(v,r)$ is of the form \eqref{eq:BranchIf}, which is a Schwarzschild--[(A)dS] black hole. The static end state is thus a GR solution of the same form as the charged black hole solutions found in \cite{Babichev:2014fka}. Setting the electric charge to zero, the end state belongs to the non-bidiagonal static, spherically symmetric vacuum solutions of \cite{Comelli:2011wq} (see also \cite{Torsello:2017ouh} for Eddington--Finkelstein coordinates).

\paragraph{Branch II.} The static end states of Branch II are,
\begin{subequations}
	\begin{alignat}{2}
		&m(v,r) = \widetilde{m}(v,r) = \mgconst + \frac{1}{6} \widetilde{\Lambda} c^2 r^3, \quad &&\mathrm{Branch \, IIa,}\\
		&m(v,r) = \widetilde{m}(v,r) = \mgconst, \quad &&\mathrm{Branch \, IIb}.
	\end{alignat}
\end{subequations}
These are proportional Schwarzschild--[(A)dS] solutions. There is no electric charge of the final black hole in this branch. As shown in section \ref{sec:EoMsol}, Branch II solutions are special as they are compatible only with a particular type of generalized Vaidya fluid. Thus, within our set of Vaidya solutions, the proportional Schwarzschild black holes are only realized as end states of gravitational collapse under special conditions. Moreover, this type of end state is potentially problematic due to a Gregory--Laflamme-like instability at the linear level \cite{Babichev:2013una}. \newpage

\subsection{Singularities and horizons}
\label{sec:singandhor}
\begin{figure}[t]
	\centering
	\includegraphics[width=1\linewidth]{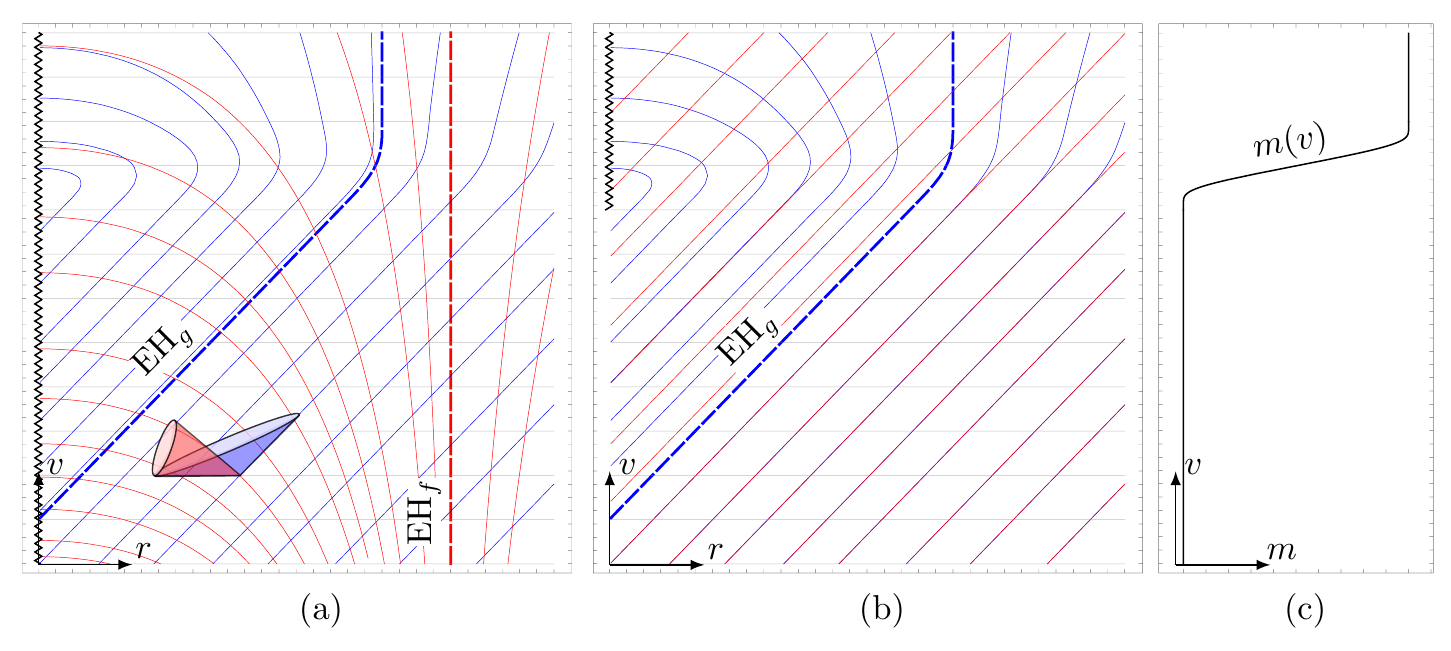}
	\caption{Branch I. (a)-(b): radial null geodesics (RNG), with the outgoing RNG of $g$ in blue and the outgoing RNG of $f$ in red. The metrics share the ingoing RNG (horizontal). The event horizons (dashed) are plotted for the two metrics. The null cones with respect to $g$ and $f$ are drawn at a common point. In the plot, $\Qv=0$, and $\Lambda = \widetilde{\Lambda}=0$, consistent with a small cosmological constant. (a) $\mfconst = 1.2 \, \mgconst$, so at $r=0$ there is a singularity due to the curvature singularity of $f$. This is only partly covered by the event horizon with respect to $g$. (b) $\mfconst=0$ and the singularity is covered by the $g$ event horizon but there is no event horizon in the $f$-sector. (c) mass, $m$, as a function of time. All plots have the same vertical axis.}
	\label{fig:br-vaidya-horizons-and-mass}
\end{figure}
\begin{figure}[t]
	\centering
	\includegraphics[width=0.7\linewidth]{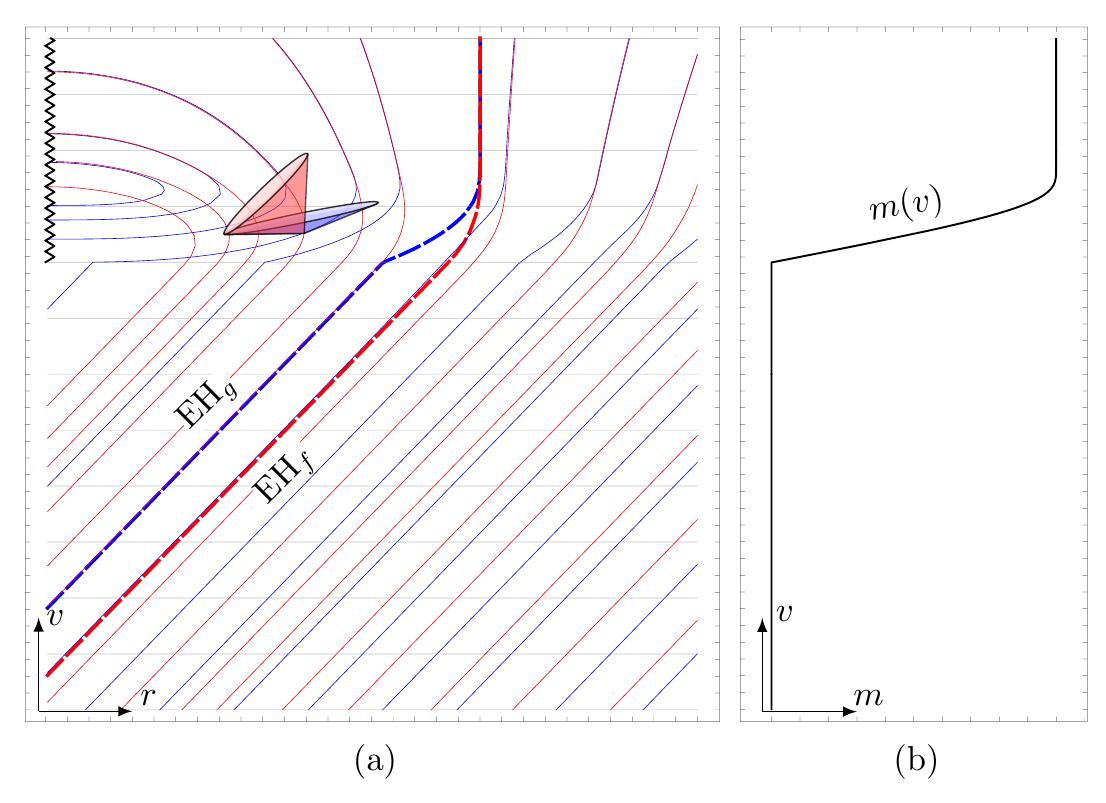}
	\caption{Branch IIa. (a) outgoing RNG of $g$ and $f$ in blue and red, respectively. The metrics share the ingoing RNG (horizontal). The event horizons (dashed) are plotted for the two metrics. The null cones of $g$ and $f$ are drawn at a common point. At $(v\geq 0,r=0)$ there is a singularity due to curvature singularities of $g$ and $f$. Both event horizons cover the singularity. (b) mass, $m$, as a function of time. In the plot, $\widetilde{\Lambda}=0$ and $\kappa_g^{-1}=c= \shiftpol{c}{1}{2}=1$, for illustrational purposes. The left and right plots have the same vertical axis.}
	\label{fig:br-vaidya-horizons-and-mass-branchii}
\end{figure}
\paragraph{Branch I.} Generically, $f$ is a Schwarzschild--[(A)dS] metric. Therefore, there is, in general, a curvature singularity with respect to $f$ at $r=0$ for all $v$. It is covered by an event horizon with respect to the same metric, located at some constant $r$. The singularity is only partially covered by the $g$-sector event horizon, see figure \ref{fig:br-vaidya-horizons-and-mass}. In this sense, a bimetric analog of the cosmic censorship hypothesis is violated for these solutions. A standard matter observer measuring the geometry of $g$ will not notice anything special when approaching the centre, but suddenly, when reaching $r=0$, the worldline ends abruptly. However, being an integration constant, the mass of the $f$-sector black hole can be set to zero, removing the naked singularity. The horizons of $g$ and $f$ do not coincide, but the theorem of \cite{Deffayet:2011rh} is not violated since the metrics are not bidiagonal. 

Concerning the square root $\Sqrt$, \eqref{eq:S}, the coordinate invariants \eqref{eq:Scoordinv} are regular within the coordinate range $0 < r < \infty$ and, moreover, in the limit $r \to 0$. However, due to the $r^{-1}$ dependence of $S^r{}_v$, $\det S$ exhibits a singularity precisely at $r=0$, unless $m -\widetilde{m} \sim r^{n\geq1}$. The only caveat is that $r=0$ is not within the allowed coordinate range. Nevertheless, transforming to Cartesian coordinates reveals the same behavior. In the initial Minkowski region of $g$ (i.e., when $v<0$), $f = S^\tr g S = S^\tr \eta S$, so $\Sqrt$ is in fact the vielbein of $f$. With $f$ being Schwarzschild--[(A)dS], the square root represents the transformation of $f$ into the local Lorentz frame and must be singular at $r=0$, due to the singularity of Schwarzschild--[(A)dS] at that point. With $f$ being (A)dS/Minkowski, there is no singularity.

Furthermore, if $f$ is Schwarzschild--[(A)dS], the Ricci scalar of the geometric mean metric, $h_{\mu\nu} := g_{\mu\rho} S^\rho{}_\nu$, diverges in the limit $r \to 0$. This supports the conjecture in \cite{meangauge}, that at least two of the three metrics $g$, $f$, and $h$ share the curvature singularity.

\paragraph{Branch II.} In the static region of Branch IIa, the two metrics, being proportional, share event horizons, as required \cite{Deffayet:2011rh}. In the intermediate Vaidya region, the $1/r$ term in the $g$-sector pulls the event horizon inwards if $\shiftpol{c}{1}{2}>0$, see figure \ref{fig:br-vaidya-horizons-and-mass-branchii}. Since the event horizon with respect to $g$ lies inside that of $f$, an observer, minimally coupled to $g$, does not see the $f$ horizon. The $1/r$ contribution is inversely proportional to $1/\ell$ \eqref{eq:BranchIIag} and hence the singularity is enclosed by both event horizons if $\ell$ is not too big. In Branch IIb, the two metrics, being proportional, share event horizons.

\section{Discussion and outlook}
\begin{comment} Given the need for bimetric numerical relativity, the bimetric Vaidya solutions may provide a testbed for numerical solutions of gravitational collapse. The ``time" coordinate of Vaidya is along a null direction, so to be able to compare with numerical results from a standard 3+1 decomposition, one has to rewrite the solution in that formalism. Another challenge is presented by the naked singularity in Branch I, which can be difficult to handle numerically. Moreover, the mass function $m(v,r)$ is basically a freely specifiable function. In particular, the profile along the coordinate, $v$, running along the common null direction, is freely specifiable. This fact may provide additional challenges when recast as an initial value problem in the standard 3+1 formalism.
\end{comment}
In bimetric theory, the two metrics share the manifold. For a global solution, one must ensure that both metrics are compatible with the topology of the manifold \cite{Torsello:2017ouh}. In Branch II, where they are Vaidya metrics, one may expect the associated topologies to concur, but in Branch I the agreement is not obvious, with $g$ being Vaidya and $f$ being Schwarzschild--[(A)dS]. If they do not agree at face value, there are several possibilities to harmonize; one obvious thing that changes the prima facie topology associated with $g$, is that we need to remove $r=0$ from our manifold due to the singularity in $f$. Another possibility is to look at the universal covering space. Also, if the space-time is regarded as an approximate model, valid in some local region, we can disregard the global structure of the manifold. Additional insight could be gained by constructing the Penrose--Carter diagram with respect to both metrics. This is left for future work. 

Besides topology, the bimetric Vaidya solutions present further questions; in Branch I, there is a naked singularity with respect to standard matter observers, unless the mass of the $f$-sector black hole is set to zero. A natural follow up question is if this is a generic property of bimetric space-times or if it is a special feature of this solution. Another important question is whether the solutions are stable against dynamical perturbations. Linearizing around a general Branch I solution is left for future work. 

In the static regions of Branch IIa (i.e., the initial state and the end state), the metrics are proportional and $c$ is the proportionality constant. Requiring $\shiftpol{c}{1}{2}>0$ in order to satisfy the energy conditions implies an imaginary Fierz--Pauli mass for perturbations around these proportional backgrounds \cite{Hassan:2012wr} (see also \cite{polytropes} for a discussion). Contrary to tachyonic particles, the causality (i.e., the causal cone of the EoM of the perturbations) is unaffected by this fact, but rather signals an instability of the background solutions \cite{Aharonov:1969vu} (however, see also \cite{Sushkov:2015fma,Breitenlohner:1982bm}). It should be stressed that such an instability does not necessarily indicate a pathology of the theory but just the fact that proportional solutions settle down to another stable end state upon perturbation. Compare for example with the Higgs mechanism in the standard model of particle physics. The cure of the instability is obviously to choose $\shiftpol{c}{1}{2}<0$ and the price to pay is that the dominant energy condition is violated outside some radius, besides the fact that it represents a radiating solution rather than gravitational collapse. For radial perturbations, instead of the ordinary Yukawa decay, there is a $1/r$ decay modulated by a sin/cos-function (in addition to the ordinary Newtonian decay), introducing oscillations around the Newtonian fall-off of the perturbations.

Even though a toy model like the one presented here may provide important insights and questions, the ultimate enigma remains: what is the end state of gravitational collapse, given a fully realistic model? To answer this question, we have to solve the bimetric equations numerically, a challenging task indeed \cite{Kocic:2018ddp,Kocic:2018yvr,cbssn,Kocic:2019zdy,polytropes,meangauge,bimEX}.

\acknowledgments
Thanks to Ingemar Bengtsson for a careful reading of a manuscript and to four anonymous referees for valuable comments. 

\appendix
\section{\change{Fierz--Pauli mass of the Branch I solutions}}
\label{sec:FPmass}
\change{In bimetric gravity, the mass of the massive graviton (the Fierz--Pauli mass) is defined by considering perturbations around proportional background solutions $f = c^2 g$,}
\change{\begin{equation}
\label{eq:mFP}
		m_\mathrm{FP}^2 = \frac{\kappa_g }{\ell^2} \left(1 + \frac{1}{\kappa c^2}\right) \left(\beta_1 c + 2 \beta_2 c^2 + \beta_3 c^3\right), \quad \kappa := \frac{\kappa_g}{\kappa_f}.
\end{equation}}	
\change{The metrics of the Branch I generalized Vaidya solutions are given by,}
\change{\begin{subequations}
		\begin{align}
		g &= -\left(1- \frac{2 m(v,r)}{r}\right) \mathrm{d}v^2 + 2 \mathrm{d}v \mathrm{d}r +r^2 \mathrm{d}\Omega^2,\\
		f &= c^2 \Bigg[-\left(1-\frac{2 \widetilde{m}(v,r)}{r}\right) \mathrm{d}v^2 + 2 \mathrm{d}v \mathrm{d}r + r^2 \mathrm{d}\Omega^2\Bigg].
		\end{align}
	\end{subequations}
}
\change{together with,}
\change{\begin{subequations}
		\begin{align}
		m(v,r) &= \mv(v) - \frac{\Qv^2(v)}{2r} + \frac{\lambda}{6}r^3  ,\\
		\widetilde{m}(v,r) &=\mfconst + \frac{\widetilde{\Lambda}}{6} c^2 r^3.
		\end{align}
\end{subequations}}
\change{The defining condition for Branch I is $\beta_1 c+ 2 \beta_2 c^2+ \beta_3 c^3=0$ \eqref{eq:cBranchI}, similar to setting the Fierz--Pauli mass to zero in \eqref{eq:mFP}. However, this is not the case since the two metrics are not proportional in general and the Fierz--Pauli mass is not defined.}

\change{Only under special assumptions, $m_\mathrm{FP}$ can be defined in the asymptotic spatial infinity ($r \to \infty$) where $m =\lambda r^3 /6$ and $\widetilde{m} = \widetilde{\Lambda} c^2 r^3 / 6$. Namely, if we assume that there are no matter fields contributing to the cosmological constant so that $\lambda = \Lambda$ and that $\Lambda = \widetilde{\Lambda} c^2$, the metrics are proportional at asymptotic infinity. The latter requirement translates to a condition on the $\beta$-parameters.\footnote{Expanding the condition, $\kappa c^2 (\beta_0 + 3 \beta_1 c + 3 \beta_2 c^2 + \beta_3 c^3) = \beta_1 c + 3 \beta_2 c^2 + 3 \beta_3 c^3 + \beta_4 c^4$. This equation is invariant under the rescaling symmetry $(f_{\mu\nu} , \kappa_f , \beta_n) \to (\omega f_{\mu\nu} , \omega \kappa_f , \omega^{-n/2} \beta_n)$ for constant $\omega$ of the bimetric action \eqref{eq:action}. Hence, the equation cannot be satisfied by choosing a specific normalization for $c$.} Even under these assumptions, it is not obvious that the solution is problematic. In contrast to homogeneous and isotropic cosmological solutions, here the Fierz--Pauli mass is vanishing at $r=\infty$ which is infinitely far from the region we are interested in (i.e., around $r=0$). In the asymptotic region, radial perturbations have the traditional $1/r$ Newtonian decay, that is, the Yukawa term is absent \cite{Comelli:2011wq}.}

\bibliographystyle{JHEP}
\bibliography{biblio}

\end{document}